\documentstyle[12pt]{article}
\font\frbig=eufb10  scaled\magstep1
\font\frscr=eufb7 scaled\magstep1
\font\frscrscr=eufb5 scaled\magstep1

\newfam\frfam
\textfont\frfam=\frbig
\scriptfont\frfam=\frscr
\scriptscriptfont\frfam=\frscrscr
\def\fr{\fam\frfam}

\font\openbig=msym10  scaled\magstep1
\font\openscr=msym7 scaled\magstep1
\font\openscrscr=msym5 scaled\magstep1

\newfam\openfam
\textfont\openfam=\openbig
\scriptfont\openfam=\openscr
\scriptscriptfont\openfam=\openscrscr
\def\open{\fam\openfam}

\font\Scbig=cmss10  scaled\magstep1
\font\Scscr=cmss8 scaled\magstep1
\font\Scscrscr=cmss8

\newfam\Scfam
\textfont\Scfam=\Scbig
\scriptfont\Scfam=\Scscr
\scriptscriptfont\Scfam=\Scscrscr
\def\Sc{\fam\Scfam}

\makeatletter

\newdimen\normalarrayskip              % skip between lines
\newdimen\minarrayskip                 % minimal skip between lines
\normalarrayskip\baselineskip
\minarrayskip\jot
\newif\ifold             \oldtrue            \def\new{\oldfalse}
\def\arraymode{\ifold\relax\else\displaystyle\fi} % mode of array enrties
\def\@arrayskip{\ifold\baselineskip\z@\lineskip\z@
     \else
     \baselineskip\minarrayskip\lineskip2\minarrayskip\fi}
\def\@arrayclassz{\ifcase \@lastchclass \@acolampacol \or
\@ampacol \or \or \or \@addamp \or
   \@acolampacol \or \@firstampfalse \@acol \fi
\edef\@preamble{\@preamble
  \ifcase \@chnum
     \hfil$\relax\arraymode\@sharp$\hfil
     \or $\relax\arraymode\@sharp$\hfil
     \or \hfil$\relax\arraymode\@sharp$\fi}}
\def\@array[#1]#2{\setbox\@arstrutbox=\hbox{\vrule
     height\arraystretch \ht\strutbox
     depth\arraystretch \dp\strutbox
     width\z@}\@mkpream{#2}\edef\@preamble{\halign \noexpand\@halignto
\bgroup \tabskip\z@ \@arstrut \@preamble \tabskip\z@ \cr}%
\let\@startpbox\@@startpbox \let\@endpbox\@@endpbox
  \if #1t\vtop \else \if#1b\vbox \else \vcenter \fi\fi
  \bgroup \let\par\relax
  \let\@sharp##\let\protect\relax
  \@arrayskip\@preamble}

\makeatother

\begin{document}
\hfuzz=1pt

\date{(September  1991)}
\title{\sc Virasoro Action and Virasoro Constraints on Integrable
 Hierarchies of the $r$-Matrix Type}
\author{A.~M.~Semikhatov\\ \sl Theory Division, P.N.Lebedev Physics Institute\\
\sl 53 Leninsky prosp., Moscow SU 117924 USSR}

\maketitle

            \begin{abstract}
For a large class of hierarchies of integrable equations admitting a
classical $r-$matrix, we propose a construction for the Virasoro algebra action
on the Lax operators which commutes with the hierarchy flows. The construction
relies on the existence of dressing transformations associated to the
$r$-matrix and does not involve the notion of a tau function. The
dressing-operator form of the
Virasoro action
gives the corresponding formulation of the Virasoro
constraints on hierarchies of the $r-$matrix type.
We apply the general construction to several examples which include
KP, Toda and generalized KdV hierarchies, the latter both in scalar and the
Drinfeld-Sokolov formalisms. We prove the consistency of Virasoro action on
the scalar and matrix (Drinfeld-Sokolov) Lax operators, and make an
observation on the difference in
the form of the string equation in the two formalisms.
            \end{abstract}

\newpage
{\Large \bf 0. Introduction}

\vspace{0.5cm}
A striking feature of the completely integrable systems, besides the
integrability by itself, is the number and diversity of their physical
applications   [!?]~\footnote{ Any
list of references appropriate for the present paper, which is not
a special review article, would be inadequate.}.
Yet another one has been discovered recently: completely integrable systems
appear in the study of non-perturbative two-dimensional quantum
gravity in the formalism of Matrix Models
 \cite{[D],[BDSS],[GMMMO],[Ma]} \footnote{ This new development was
 initiated by refs.\cite{[BK],[DSh],[GM]}; for a review see, for
 instance, \cite{[Ag]} and references therein.  } .

Integrable equations turn out to govern exact ``renormalization-group''
evolutions in the space of coupling constants. As there are infinitely many
coupling constants in gravitational theories, it is perhaps less surprising
that one actually obtains {\it hierarchies} of integrable equations.

The fact that really came about as a surprise, that {\it integrable} equations
appear at all, might seem less unexpected if one notes that orthogonal
polynomials, which are a standard technique to work with matrix integrals
\cite{[BIZ]}\cite{[IZ]}, do in a sense {\it solve} non-linear integrable
equations.

Further, solutions to integrable hierarchies that do come from a matrix model
are by no means general (nor of much independent interest to the experts in
non-linear integrable equations!). The special class of solutions related
to matrix models is singled out by the highest-weight conditions on the
tau function with respect to a certain Virasoro algebra
\cite{[FKN1],[DVV],[MM],[IM1]}.  These conditions are usually
called the Virasoro constraints. They were first recognized in the apparently
weaker guise of the string equation $\cite{[D]}$, which is in fact a
consequence
of the vanishing condition under the $L_{-1}$ Virasoro generator. For a number
of cases, the machinery of integrable systems allows one to prove all the
$L_{\geq 0}$ vanishing conditions starting from the string equation.

The form of the string equation, however, is tied up with particular properties
of the linear system associated to the hierarchy. The Virasoro constraints, on
the other hand, appear to have a more universal meaning, as they are of
essentially the same form for a wide class of integrable hierarchies. Thus,
despite the success of the use of the string equation in applications,
\cite{[DfK],[GGPZj]},  we prefer to promote the Virasoro constraints to a
"first-principle" of a hierarchial description of gravity-coupled
two-dimensional theories. More generally, one should not be limited to
the {\it Virasoro} constraints alone: Much as in the historical development of
conformal field theory in two dimensions, higher symmetry algebras extending
Virasoro have been identified in the analysis of matrix models, namely the
$W-$algebras \cite{[FKN1],[FKN2],[G],[AS],[Y],[IM2]}.
Expected next might be highest-weight conditions with respect to a semi-direct
product of Virasoro and Kac-Moody algebras. Such a possibility for the $N$-KdV
hierarchies was pointed out in \cite{[S6]}.

Thus, let us assume that the constraints with respect to some good algebra
imposed on integrable hierarchies may serve as a counterpart of the
field-theoretic description. Then it is not only the algebra but also the
hierarchy itself, that is optional, and our aim in this paper will be to
investigate Virasoro constraints on integrable hierarchies of as general form
as possible. We will in fact propose a rather general construction for the
Virasoro algebra {\it action} on integrable hierarchies and then apply it to
the study of the constraints as the invariance conditions under this action.

The formalism which we are going to use, and which allows considerable
generality, is that of the classical $r-$matrix \cite{[Sts1]}\cite{[Sts2]} and
dressing
transformations related to the $r-$matrix. Integrability of essentially
all integrable systems can be 'explained' by an underlying $r-$matrix.
Our starting point will thus be an abstract $r-$matrix satisfying the
classical Yang-Baxter equation.

In contrast with most of the recent papers dealing with Virasoro constraints in
Matrix Models, we do not rely on the existence of a tau function. The Virasoro
algebra will be represented in our approach on the Lax operators of an abstract
hierarchy, $i.e.$, on certain coadjoint orbits. By specializing our general
construction to several popular examples we will be able to reproduce the known
results on the Virasoro algebra action.

We therefore start in Sect.1 with defining hierarchies of the $r-$matrix
type. Further, in Sect. 2, we specialize to those hierarchies which admit a
Virasoro algebra action. This, of course, requires making additional
assumptions on the ingredients of the $r-$matrix formalism, so we proceed
with first stating "kinematical" axioms and then translate them into a
construction of the Virasoro action compatible with the equations of a
hierarchy.

In sections 3 and 4 we consider a number of examples of the general
construction. The KP case, chosen for its special simplicity, is purely
illustrative, while the relation between Virasoro constraints and string
equations in two different formalisms for generalized KdV-hierarchies
reveals certain subtleties and may be of an independent interest to the
experts.

\section{ Classical $r-$matrices and integrable hierarchies}

\leavevmode\hbox to\parindent{\hfill}
In this section we recall a number of basic facts about classical
$r-$matrices and their use for the construction of integrable systems.
The readers familiar with refs.\cite{[Sts1]}\cite{[Sts2]} may skip to the next
section
(however, we borrow the presentation partly from \cite{[KsM]}).

\vspace{0.3cm}
Let $( {\fr g}, [~,~])$ be a Lie algebra and $ {\fr g}^\ast $ its
dual space.

\vspace{0.3cm}
{\bf {\it 1.1.}}{\sc Definition}. The {\sl classical $r-$matrix} is a linear
mapping $$ r: { \fr g}^\ast  \rightarrow   {\fr g}
$$ such that the {\sl classical Yang-Baxter equation} (CYBE)
$$ r({\rm ad}^\ast _{^tr\xi }.\eta  + {\rm ad}^\ast _{r\eta }.\xi ) + [r\xi ,
r\eta ] =0
$$ is satisfied for all $\xi , \eta  \in   {\fr g}^\ast $. Here ${}^tr$
denotes the transposed mapping ${}^tr: { \fr g}^\ast  \rightarrow   {\fr g}.$

\vspace{0.3cm}
{\bf {\it 1.1.1.}} {\sc Remark}. To make contact with the usual 'tensor'
formulation of the classical Yang-Baxter equation, one uses the isomorphism
$$
{\rm Hom}({\fr g}^\ast , { \fr g}) = { \fr g} \otimes  { \fr g}
$$
to identify $r$  with a tensor
$$
r = r^{ab}t_a \otimes  t_b \in  {\fr g} \otimes  {\fr g}.
$$
With the components of $r$ w.r.t. a basis $\{t_a\}$ in  ${\fr g}$ thus
defined, we have
$$
r\xi  = \xi _a r^{ab} t_b,\quad     {}^tr = \xi _a r^{ba} t_{b}.
$$
Now contracting the {\sc lhs} of the CYBE, which is an element of ${ \fr g}$,
with a $\zeta  \in  { \fr g}^\ast $, we rewrite the result as, $$
\new \begin{array}{rcl}
{}&-& \langle  \eta , [r\xi , r\zeta ] \rangle  - \langle \xi , [{}^tr\eta ,
r\zeta ] \rangle  + \langle  \zeta , [{}^tr\xi ,  {}^tr\eta ] \rangle \\
{}&=& \langle  \zeta \otimes  \xi \otimes  \eta , [r_{13}, r_{23}] + [r_{12},
r_{23}] + [r_{12}, r_{13}] \rangle \end{array}
$$ where
$$ r_{12} = r^{ab} t_a \otimes  t_b \otimes  1,\qquad    r_{13} = r^{ab} t_a
\otimes  1 \otimes  t_b,\qquad   r_{23} = r^{ab} 1 \otimes  t_a \otimes t_b.
$$ Thus we arrive at the standard form
$$ [r_{13}, r_{23}] + [r_{12}, r_{23}] + [r_{12}, r_{13}]
= 0 $$

\vspace{0.3cm}
{\bf {\it 1.2.}} Decomposing $r$  into its symmetric and antisymmetric parts,
$$
s = {1\over 2} (r + {}^tr),
 \qquad a = {1\over 2} (r - {}^tr),
$$
{\sl assume that $s$ is invertible and ad-invariant}:
$$
s \circ  {\rm ad}^\ast _x = {\rm ad}_x \circ s\quad {\rm   for all }\quad x \in
{\fr g}.  $$
(We thus invoke a Killing form on ${\fr g}$, which is the case  encountered in
applications.)
Then all the "degrees of
freedom" of an $r-$matrix are contained in its antisymmetric part.

{\bf {\it1.3.}} With the above assumptions on $s$, introduce
$$
R = a \circ  s^{-1}: { {{\fr g}}} \rightarrow  { {{\fr g}}}
$$
Then, by virtue of 1.1,
$$
[ Rx, Ry ] - R([Rx, y ] + [x,  Ry ]) = - [x,y ]
$$
holds for any $x,y \in  {\fr g}$. This will be called the {\sl modified
classical Yang-Baxter equation} (mCYBE).

{\bf 1.4.} As a straightforward consequence of the mCYBE note that
$$
{\fr g}_- \equiv  {1\over 2} (1 - R){\fr g}
$$
is a subalgebra in ${\fr g}$. For any $x \in  {\fr g}$ we denote
$$
x_- = {1\over 2} (1 - R)x
$$
The mapping  ${1\over 2}(1 - R): {\fr g} \rightarrow  {\fr g}_-$
provides us with an
embedding ${\fr g}^\ast_- \rightarrow {\fr g}^\ast $. We will identify
${\fr g}^\ast _-$ with its image in ${\fr g}^\ast $. Then for an $m \in  { \fr
g}_-$ one should clearly distinguish between the representation $$ {\rm
ad}^\ast
_m: {\fr g}^\ast  \rightarrow  {\fr g}^\ast $$ and the properly coadjoint
action
$$ {\rm ad}^\ast _{(-)m}: {\fr g}^\ast _- \rightarrow  {\fr g}^\ast _- $$ of
the
${\fr g}_-$ algebra on its dual space.

Define, formally, $G_- = \exp \ {\fr g}_-$ to be the Lie group
of ${\fr g}_-$. We will
call it the group of {\sl dressing transformations} and its elements,
{\sl dressing operators}.

\vspace{0.3cm}
{\bf {\it 1.5.}} Next, let us fix  an element
$$
\Lambda  \in  { \fr g}^\ast _-
$$
such that
$$
{\rm ad}^\ast _{(-){ \fr g}^{}_-}.\Lambda  = 0.
$$
Viewing $\Lambda $ as an element of ${ \fr g}^\ast $, let
${\cal Z}_\Lambda$ be the maximum commutative subalgebra of
the invariance subalgebra of $\Lambda $ in ${ \fr g}$, i.e., of
$$
\left\{  x \in  { \fr g} \mid
  {\rm ad}^\ast _x.\Lambda  = 0
\right\}
$$
(In applications to the Kac-Moody algebras, for instance, this
invariance subalgebra would by itself be commutative for {\it regular}
$\Lambda$.)

{\bf {\it 1.6.}} The elements of the ${\rm ad}^\ast$-orbit of $G_-$ in ${\fr
g}^\ast $ going through $\Lambda $, will be called Lax operators $Q$~:
$$
Q \equiv  Q(K) = {\rm Ad}^\ast _K .\Lambda  \in  { \fr g}^\ast
\qquad   K \in  G_- $$

{\bf {\it 1.7}}. For any $a \in  {\cal Z}_\Lambda $ we introduce an evolution
on
the {\rm Ad}- and ${\rm Ad}^\ast -$orbits of $G_- $ \footnote{ Some of the
evolution flows may happen to be `trivial' (as are, for instance, {\it linear}
equations). Fewer such `trivial' equations are among those associated to $a \in
{\cal Z}_\Lambda  \cap  { \fr g}_+$. We will ignore this point, however.
} . First, for ${\cal X} = {\rm Ad}_K .x$ we set
$$
{\partial \over \partial t} {\cal X} = - [{\cal A}_-,
{\cal X} ]
$$
where
$$
{\cal A} \equiv  {\cal A}(K) = {\rm Ad}_K .a $$
Similarly, for $\Xi  = {\rm Ad}^\ast _K .\xi $ being an element of the
coadjoint
orbit,
$$
{\partial \over \partial t} \Xi  = -{\rm ad}^\ast _{{\cal A}_-}.\Xi
$$

{\bf {\it 1.8}}. A generic evolution equation of the hierarchy that we will
associate to the triple $( { \fr g}, r, \Lambda )$ is the ${\rm ad}^\ast -$
equation of 1.7 imposed on the Lax operator $Q $:
$$
{\partial \over \partial t} Q = - {\rm ad}^\ast _{{\cal A}_-}.Q
$$
$$
Q = {\rm Ad}^\ast _K .\Lambda \qquad    {\cal A} = {\rm Ad}_K
.a, \qquad    a \in {\cal Z}_\Lambda .
$$

{\bf {\it 1.9}}. {\sc Lemma}. {\sl Any two flows of the type 1.8 commute.}

\noindent
Indeed, consider the flows
$$
{\partial \over \partial t_i}Q =
-{\rm ad}^\ast _{{1\over2}(1-R){\cal A}_i}.Q, \qquad  i = 1, 2.
$$
with
$$
{\cal A}_i = {\rm Ad}_K .a_i\qquad    a_i \in  {\cal Z}_\Lambda .
$$

\noindent
Then, using 1.7 and 1.8, evaluate
$$
{\partial \over \partial t_1} {\partial \over \partial t_2} Q =
{\rm ad}^\ast _{{1\over2}(1-R)[{1\over2}(1-R){{\cal A}_2} ,  {\cal A}_1]}.Q +
{\rm ad}^\ast _{{1\over2}(1-R){\cal A}_1}.{\rm ad}^\ast _{{1\over2}(1-R){\cal
A}_2}.Q
$$
whence the antisymmetrized second derivative equals,
$$
\left( {\partial \over \partial t_1} {\partial \over \partial t_2} -
{\partial \over \partial t_2} {\partial \over \partial t_1}\right) Q =
{\rm ad}^\ast _{\omega (1,2)}.Q
$$
with
$$
 \new
\begin{array}{rcl}
\omega (1,2) &=& - {1\over 4} (1 - R)[(1 - R){\cal
A}_1, {\cal A}_2 ] - {1\over 4} (1 - R)[{\cal A}_1, (1
- R){\cal A}_2 ] + {1\over 4} [(1 - R){\cal A}_1, (1 - R){\cal A}_2
]\\
{} &=& - {1\over 2} (1 - R) [{\cal A}_1, {\cal A}_2 ] \hfill {\rm (by \, \,
1.4)}\\

{} &=& - {1\over 2} (1 - R) {\rm Ad}_K .[a_1, a_2 ]\\

{} &=& 0        \hfill             {\rm (by  \, \, 1.5)}
\end{array}
$$
$\Box$

\vspace{0.3cm}
 Thus, picking up different elements of $ {\cal Z}_\Lambda $, which
in applications is usually in\-fi\-nite-di\-men\-sional, one can generate an
(infinite) set of mutually commuting differential equations which we call the
 hierarchy associated to $( { \fr g}, r, \Lambda )$.

\vspace{0.3cm}
At the closing of this section let us present a construction which will be used
below to implement {\it symmetries} of hierarchies of the above type.

\vspace{0.3cm}
 {\bf {\it 1.10}}. Every $ v \in  { \fr g}$
 defines a vector field $ \hat v$ on {\rm Ad}- and
 ${\rm Ad}^\ast$-orbits of $G_-$. That is, for a function $f~:~{\rm
 Ad}_{G_-}.x \rightarrow  {\open C}$, the action of $\hat v$ on $f$  is defined
by
$$
(\hat vf)({\cal X}) = {d\over d\epsilon } f({\cal X} -
\epsilon  {\rm ad}_{{\cal V}_-}.{\cal X} )\mid_{\epsilon =0},
$$
$$
{\cal X} = {\rm Ad}_K .x\qquad   {\cal V} = {\rm Ad}_K .v
$$
(recall that $(\ldots )_-$ was defined in 1.4). Similarly, for a function
 $ \phi : {\rm Ad}_{G_-}.\xi  \rightarrow  {\bf \open C}$,  we set
$$
(\hat v\phi )(\Xi ) = {d\over d\epsilon } \phi (\Xi  -
\epsilon \ {\rm ad}^\ast _{{\cal V}_-}.\Xi  )
\mid_{\epsilon =0},
$$
$$
\Xi  = {\rm Ad}^\ast _K .\xi .
$$

\vspace{0.3cm}
We will denote by [[~,~]] the commutator of such vector fields.

\vspace{0.3cm}
{\bf {\it 1.11}}. {\sc Lemma}. {\sl The mapping}
$$
v \mapsto  \hat v
$$
{\sl is a representation of  ${ \fr g}$ in the $[[~,~]]$-algebra
of the above vector fields}.

\noindent
{\sc Proof} follows directly from the mCYBE; the calculation is
identical to that of 1.9. $\Box$ .

\vspace{0.3cm}
Now we extend the construction of 1.10 to {\it time-dependent} $v$
that deform the Lax operators.

\vspace{0.3cm}
{\bf {\it 1.12}}. {\sc Lemma}. {\sl The condition for a vector field $ \hat v$
to
commute with the hierarchy flows 1.8, $i.e.$, with

$$
{\partial \over \partial t_a}Q = -
{\rm ad}^\ast _{{1\over2}(1-R){\rm Ad}_K .a}
.Q , \qquad a \in  {\cal Z}_\Lambda ,
$$
is
$$
{\rm ad}^\ast _{{\cal F}_-}.Q = 0
$$
where}
$$
{\cal F} \equiv  {\rm Ad}_K .\left( {\partial v\over \partial t_a} +
[v,a]\right) .
$$

\noindent
{\sc Proof}. Applying $\hat v$ to deform the equation of motion, we find
$$
\delta  {\partial Q\over \partial t} =
{\rm ad}^\ast _{{1\over2}(1-R)[{1\over2}(1-R){\cal V}, {\cal A} ]}.Q -
{\rm ad}^\ast _{{1\over2}(1-R){\cal A}} .{\rm ad}^\ast _{{1\over2}(1-R){\cal
V}} .Q
$$
where
$$
{\cal V} \equiv  {\rm Ad}_K .v, \qquad   {\cal A} \equiv  {\rm
Ad}_K .a
$$
On the other hand, we use 1.7 to evaluate ${\partial \over
\partial t}\delta Q$ along the evolution flow. Then in the difference
$[\delta , {\partial \over \partial t} ]Q$ we again use the mCYBE, and thus
demonstrate that it indeed equals  $- {1\over 2} {\rm ad}_{{\cal F}_-}.Q$ .
$\Box$ .

{\bf {\it 1.13}}. We will actually use the sufficient condition
$$
{\partial v\over \partial t_a} + [v,a] = 0
$$
for a time-dependent vector field $\hat v$ to commute with a hierarchy
equation.

\section{Hierarchies with the Virasoro algebra action}
\leavevmode\hbox to\parindent{\hfill}
Now we specify the construction of the previous section to hierarchies
which admit a Virasoro algebra action. This will require adopting certain
further postulates concerning the triple $( { \fr g}, r, \Lambda )$: for
instance, the Lie algebra  ${ \fr g}$ itself has to be 'big' enough for the
theory based on it to admit a Virasoro action at all. We start in this section
with rather mild assumptions on $( { \fr g}, r, \Lambda )$, and the
Virasoro algebra will not appear until 2.5.

\vspace{0.3cm}
The first extension of the data we had in Sect.1 is achieved by allowing $v$
from 1.10 to take values in the derivation algebra
 ${\rm Der} { \fr g}$ of ${ \fr g}$.

\vspace{0.3cm}
{\bf {\it 2.1}}. Extend the adjoint action ${\rm Ad}_G$ to ${\rm Der}{ \fr g}$
by
$$
{\rm Ad}_g.M = {\rm Ad}_g \circ  M \circ  {\rm Ad}_{g^{-1}} M \in  {\rm
Der}{ \fr g}
$$
For $M = {\rm ad}_m$ being an inner derivation, this reduces to the ${\rm
Ad}_g$
action on $m$.

\vspace{0.3cm}
Now we start increasing our demands on the objects we are dealing with.

\vspace{0.3cm}
{\bf {\it 2.2.}} {\sl Assume the mapping} $R : { \fr g} \rightarrow  { \fr g}$
defined in 1.3 {\sl extends to a linear mapping
$$
{\Sc R} : {\rm Der} { \fr g}
\rightarrow  {\rm Der}{ \fr g}
$$
such that
$$
[{\Sc R} M, {\Sc R}  N ] - {\Sc
R} ([{\Sc R} M, N ] + [M,{\Sc R}  N ]) = - [M, N ]
$$
for any $M,N \in  {\rm Der}{ \fr g}$, and
$$
{\Sc R}{\rm ad}_x = {\rm ad}_{Rx}
$$
for inner derivations}.

{\bf {\it 2.3.}} {\sc Comment}. For $M$ and $N$ both inner, the above equation
will be satisfied by virtue of the mCYBE. When only $N = {\rm ad}_n$ is inner,
the required relation takes the form of the condition that for any $M \in  {\rm
Der}{ \fr g}$, the mapping
$$
[({\Sc R} M),R ] - R \circ  M \circ  R + M \in  {\rm End}{ \fr g}
$$
maps  ${ \fr g}$ into its center. However, an attempt for a characterization of
additional data responsible for the desired extension of $R$ to ${\Sc R}$  does
not seem to make much sense unless a considerable extra input  is introduced
into our formal treatment (note, in particular, that no assumptions have been
made on ${\rm Der}{ \fr g}/{\rm Inn}{ \fr g}$).

On the other hand, recall that outer derivations of a Lie
algebra ${ \fr g}$ may
be thought of as induced by inner derivations of a `bigger' Lie algebra which
contains $ { \fr g}$ as an ideal. Formally, this bigger algebra is nothing but
a semidirect product of ${ \fr g}$ with ${\rm Der}{ \fr g}$. In some
cases this formal
construction happens to be itself a `good' Lie algebra, and so
all we would need in that case is an $r-$matrix defined on this bigger algebra.

{\bf {\it 2.4.}} Given an extension 2.2, one can further generalize the
construction of 1.10 to include derivations $M \in  {\rm Der}{ \fr g}$ instead
of $v \in  { \fr g}$. Then 1.11 would still hold with $M \mapsto  \hat M$
providing a representation of ${\rm Der}{ \fr g}$ in the vector fields
tangent to the orbits:
$$
(\hat Mf )({\cal X}) = {d\over d\epsilon } f({\cal X}
- \epsilon  ({\rm Ad}_K .M)_-.{\cal X} )
\mid _{\epsilon =0}.
$$
In particular, such an $\hat M$ can be used to deform the Lax operator $Q$,
and,
moreover, time-dependent $M$'s can be allowed. Then the sufficient condition
for
the compatibility of $\hat M(t)$ with the hierarchy equations would read
$$
{\partial M\over \partial t_a} + {\rm ad}_{M.a} = 0
$$
{\sl Thus only the time-independent part of $M$ can be an outer derivation}.
However, for notational convenience, we will mostly write $M = {\rm ad}_m$ even
when $M \notin  {\rm Inn}{ \fr g}$ and thus $m \notin  { \fr g}$. An
appropriate
formal setting would again be the semidirect product of  ${\fr g}$ with ${\rm
Der} { \fr g}$, but we will not make it explicit. Accordingly, by a slight
abuse of notation, we will find it convenient to denote the above vector field
as $\hat m$ rather than $\hat M.$

\vspace{0.3cm}
Now we narrow the class of the triples $({ \fr g}, r, \Lambda )$ by
introducing the Virasoro algebra into the game. To start with, a 'kinematical'
Virasoro action will be postulated in the form of a Virasoro representation on
${ \fr g}$, while the main point of the construction to follow will consist in
dressing this action, {\it i.e.}, carrying it over to coadjoint orbits in a way
compatible with the hierarchy equations.

\vspace{0.3cm}
{\bf {\it 2.5.}} {\sl Let there be given a homomorphism {\rm Vir} $\rightarrow
{\rm Der}{ \fr g}$ of the Virasoro algebra into the derivation algebra of
${\fr g}$}. Let ${\rm ad}_{l_n }$ denote the image of the standard Virasoro
generators under this homomorphism.  Thus,
$$
[{\rm ad}_{ l_m}, {\rm ad}_{l_n }]
= (m - n) {\rm ad}_{l_{m+n}}.
$$

{\bf {\it 2.5.1.}}  {\sc Remark}. Note that ${ \fr g}$
itself is not required to be big enough to {\it contain} the Virasoro algebra
as
a subalgebra. See 2.4 concerning our notations: we do {\it not} require ${\rm
ad}_{l_j}$ to be inner derivations! Let us note once again (so as not to
return to it any more) that the $l_n$ may be viewed quite rigorously as
elements $(0, l_n)$ of the semidirect product of  ${ \fr g}$ with {\rm Vir}.

{\bf {\it 2.6}}. Further, {\sl assume that the representation of {\rm Vir}
on ${ \fr g}$
by derivations restricts to} ${\cal Z}_\Lambda $. Then we define the
derivations ${\rm ad}_{{\fr l}_j}$ with
$$
{ {{\fr l}}}_j = l_j + \sum _{a\in {\cal I}}t_a {\rm ad}_{{\fr l}^{}_j}.a
$$
where ${\cal I}$ is a set of (linearly-independent) elements from ${\cal
Z}_\Lambda $ and the times $t_a$ label the flows associated to these
elements $a$. These ${ \fr l}_j$ are now used  to construct vector
fields ${\hat { \fr l}}_j$ ,
$$
(\hat {\fr l}_j \phi )(\Xi ) = {d\over d\epsilon } \phi (\Xi  - \epsilon
{1\over 2} \left( \left( 1 - {\Sc R} )\left( {\rm Ad}_K .{\rm ad}_{{\fr
l}_j}\right) \right) .\Xi\right) \mid _{\epsilon =0},\quad   \Xi  = {\rm
Ad}^\ast _K .\xi ,\quad  \xi  \in { \fr g}^\ast .
$$
(with ${\rm Ad}_K$ defined in 2.1).

{\bf {\it 2.7}.} {\sc Theorem}.

{\bf 1}. {\sl Vector fields $\hat { \fr l}_j$, $j \in  {\open  Z}$,
furnish a representation of the Virasoro algebra on the coadjoint orbit ${\rm
Ad}^\ast_{G_-}.\Lambda$ and are compatible with the hierarchy flows associated
to $a \in  {\cal I}$, {\it i.e.,} with the flows}
$$
{\partial Q\over \partial t_a} =
- {\rm ad}^\ast_{   { {\rm Ad}^{}_K .a }}.Q,
\qquad a \in  {\cal I}
$$

{\bf 2}. {\sl Conversely, for fixed $l_j$ and ${\cal I} \subset  {\cal
Z}_\Lambda$ \footnote{that is, for a fixed homomorphism  ${\rm Vir} \rightarrow
{ \fr g}$ and a fixed subhierarchy that one wishes to consider.}, the above ${
\fr l}_j$'s are the only time-dependent derivatives ${{ \fr l}_j}(t)$ on ${ \fr
g}$ such that:

(i) $[{{ \fr l}_i}(t) ,{{ \fr l}_j}(t)] = (i - j){{ \fr l}_{i+j}}(t);$

(ii) $\hat {\fr l}_i(t)$ act as symmetries of the subhierarchy associated to}
${\cal I} \subset  {\cal Z}_\Lambda ;$

(iii) ${ \fr l}_i(0) = l_j$ .

{\sc Proof} is immediate in view of the above lemmas. 1.11 (together with 2.1
and 2.2, as discussed in 2.4) reduces the commutators $[[\hat { \fr l}_i ,\hat
{
\fr l}_j ]]$  to the 'bare' commutators $[{ \fr l}_i ,{ \fr l}_j ]$.
Compatibility with the hierarchy equations follows from 1.13 (extended to {\rm
Der}${ \fr g}$ as explained in 2.4): the ${ \fr l}$'s were constructed so
that, indeed,
$$
{\partial {\fr l}_j\over \partial t_a} + {\rm ad}_{{\fr l}_j}.a = 0
$$

Conversely, {\it solving} the equation of 2.4 for the dependence on $t_a$, we
find
$$
{{ \fr l}_i}(t) = \left( \exp \sum_{a\in {\cal I}} t_a {\rm ad}_a\right)
{{\fr l}_i}(0)
$$
which gives the above ${{ \fr l}_i}(t)$ in view of 1.5. $\Box$.

{\bf {\it 2.8.}} Later on we will encounter the following specialization of
2.7:
suppose there exist certain elements $A_m \in  {\cal Z}_\Lambda $, $m \in
{\open Z}\backslash N{\open Z}$, such that
$$
{\rm ad}_{l_m}.A_n = - {n\over N} A_{n+Nm}.
$$
Then the time-dependent Virasoro generators take the form
$$
{\fr l}_j = l_j + \sum ^N_{\alpha =1}\sum ^K_{i=0}\left( i + {\alpha \over N}
\right) t_{Ni+\alpha } A_{N(i+j)+\alpha },\qquad   j \in  {\open Z}
$$
which involves the hierarchy times $t_{Ni+\alpha }$ with $i \leq  K$ and
$\alpha  \in  \{1$, 2,\ldots , $N-1\}$, thus guaranteeing the consistency with
the
corresponding hierarchy flows. Note that to achieve consistency with {\it all}
the flows one has to make $K$ infinitely large, thus allowing infinite sums of
elements of ${ \fr g}$.

Also, given the $l_j$'s in this particular case, one can redefine them as
$$
l_j \mapsto l_j + \beta (j)A_{Nj}
$$
where the Virasoro commutation relations would require
$$
j\beta (j) - k\beta (k) = (j - k)\beta (j + k)
$$
or,
$$
\beta (j) = - j\beta (-1) + (j + 1)\beta (0)
$$
which we rewrite by introducing constants $J$ and $q$, as
$$
l_j(J,q) = l_j + (Jj + q)A_{Nj}
$$
These derivations of  ${\fr g}$ still obey the Virasoro algebra and can be used
instead of the $l_j$ to construct time-dependent derivations ${ \fr l}_j$ .

\vspace{0.3cm}
{\bf {\it 2.9.}} {\sc Remarks}.
{\bf 1.} Let us point out a heuristic construction for the $l_j$ and $A_n$,
which, however, works in a number of particular cases. First, find a derivation
${\rm ad}_p \in  {\rm Der}{ \fr g}$ such that
$$
{\rm ad}^\ast _{-p}.\Lambda  = \Lambda .
$$
Again, ${\rm ad}_p$ is not necessarily an inner derivation. Second, by the use
of the
Killing form, $\Lambda $ can be considered as an element of ${ \fr g}$, and,
further, its powers $\Lambda ^n \in  {\cal U}{ \fr g}$ may (and do in many
cases)
happen to lie in fact in ${ \fr g}$ at least for certain values of $n$ (such as
$n \in  {\open Z}\backslash N{\open Z})$. Finally, using an appropriate
associative structure, one can construct the $l_m$ as
$$
l_m = {1\over N} p\Lambda ^{Nm}.
$$

{\bf 2.} To continue the previous remark, note that the choice of $p$ is by no
means unique: the above requirements may only determine $p$  modulo
${\cal Z}_\Lambda $. A particular choice of $p$ is the choice of a
representative in ${\rm Der}{ \fr g}/{\rm ad}_{{\cal Z}_\Lambda }$.
However, we will not discuss this issue in any detail in the
present paper, as it would require introducing more special assumptions.

\vspace{0.3cm}
We finally recall that one of our motivations was to build a formalism for
dealing with the Virasoro {\it constraints} on integrable hierarchies.

\vspace{0.3cm}
{\bf {\it 2.10.}} The highest-weight conditions with respect to the Virasoro
representation constructed, take the form
$$
{ \fr L}_j \equiv  \left( {\rm Ad}_K .\left( l_j + \sum _{a\in {\cal I}}t_a
{\rm ad}_{l_j}.a\right) \right)_{!\ -} = 0,\qquad   j \geq  0.
$$\
For the reader who prefers an abuse of his patience to the abuse of notations,
we rewrite the constraints more rigorously as
$$
{\fr L}_j \equiv  {1\over 2} (1 - {\Sc R} )\left( {\rm Ad}_K .\left( {\rm
ad}_{l_j} + \sum _{a\in {\cal I}}t_a [{\rm ad}_{l_j} , {\rm ad}_a] \right)
 \right)  = 0, \qquad   j \geq  0
$$
with ${\rm Ad}_K$ defined in 2.1 and only ${\rm ad}_a$ being in general an
inner derivation
of  ${ \fr g}$.

These constraints can be imposed "off-shell", $i.e.$, independently of
the hierarchy equations, just as constraints on the dressing operator $K \in
G_-$. They in an obvious way carry a dependence on the $r-$matrix chosen,
as well as the Lie algebra ${ \fr g}$.

In particular, in the case discussed in 2.8 we find the Virasoro constraints of
the form
$$
{ \fr L}_j \equiv  \left( {\rm Ad}_K .\left( l_j + \sum ^N_{\alpha =1}\sum
^K_{i=0}( i + {\alpha \over N} )t_{Ni+\alpha } A_{N(i+j)+\alpha }\right)
\right)_{\! -}= 0,\quad   j \geq  0.
$$
Recall that these constraints are consistent with only a finite subset of
equations of the hierarchy associated to $({ \fr g}, r, \Lambda )$.
Below, in Sect.3, we will formally allow $K$ to be infinitely large, which will
allow us to achieve consistency with {\it all} the equations of certain
hierarchies.

\section{Virasoro generators and Virasoro constraints on the KP,
$N$-KdV and  Toda hierarchies}

\leavevmode\hbox to\parindent{\hfill}
{\large {\bf KP}}
\vspace{0.3cm}

{\bf {\it 3.1.}} For the KP hierarchy \cite{[DDKM]} the Lie algebra ${ \fr g} =
\psi {\rm Diff}$ is that of arbitrary order pseudodifferential operators in the
derivation $D = {\partial / \partial x}$ :
 $$
{ \fr g} = \psi {\rm Diff} \ni
\left\{ F = \sum ^n_{i=-\infty }f_i D^i \right\}
 $$
where $f_i$ are (scalar) functions of $x$.

{\bf {\it 3.2.}} The standard $r-$matrix is of the form
$$
RF = \left( \sum _{i\geq 0}f_i D^i - \sum _{i<0}f_i D^i\right) .
$$
Then
$$
F_- \equiv  {1\over 2} (1 - R)F = \sum _{i<0}f_i D^i
$$
and $G_-$ is the group of operators of the form
$$
K = 1 + \sum _{n\geq 1}w_n D^{-n},
$$
with its Lie algebra ${\fr g}_-$ being the algebra of all negative order
pseudodifferential operators.

{\bf {\it 3.3.}} The trace functional \cite{[A]} allows us to identify
${\fr g}^\ast _-$ with ${\fr g}_+$, which is the algebra of all differential
operators.

{\bf {\it 3.4}.} Let us choose
$$
\Lambda  = D \in  { \fr g}_+ \approx  { \fr g}^\ast _-
$$
For this, indeed (see 1.5)
$$
{\rm Ad}_{(-)K} .D = (KDK^{-1})_+ = D
$$
The hierarchy equations read,
$$
{\partial Q\over \partial t_n} = - [(Q^n)_- , Q],\qquad  Q =
KDK^{-1},\quad n \geq  1.
$$

{\bf {\it 3.5.}} The Virasoro generators $l_n$ admit a construction in the
spirit of 2.9:
$$ p = xD, \qquad [-p,D ] = D $$
$$
 l_n = pD^n =
xD^{n+1}
$$
and, further,
$$
A_n = D^n,
$$
so that the $(J,q)$-dependent generators of 2.8 become,
$$
{\fr l}_n(J,q) = xD^{n+1} + JnD^n + qD^n + \sum _{m\geq 1}t^{}_m m\ D^{m+n}
$$

Finally, the extensions, required in 2.1 and 2.2, of the
${\rm Ad}_{G_-}$-action  on all the derivations involved, are achieved most
straightforwardly, and so the vector fields $\hat l_n(J,q)$ act on the Lax
operator $Q$ via
$$
{\hat { \fr l}_n}(J,q).Q = - [{ \fr L}_n (J,q), Q]
$$
with
$$
{ \fr L}_n (J,q) = \left( K\left( xD^{n+1} + JnD^n + qD^n + \sum _{m\geq 1}t_m
m\ D^{m+n}\right) K^{-1}\right) _{\! -}.
$$
This form of Virasoro generators on the KP hierarchy was first arrived at in
\cite{[S1]} by a direct calculation from the well-known expression for the
Virasoro generators acting on the tau function\footnote{ In
a more recent paper \cite{[FKN3]} these generators were rederived and also
reinterpreted from the infinite-Grassmannian point of view.}.
A geometric interpretation of $J$ and $q$ was also given there.

{\bf {\it 3.6.}} There exists, however, a yet more general expression for $l_n$
and therefore for the above  ${ \fr L}$'s. To arrive at it in a systematic way,
recall that the representation $Q = KDK^{-1}$ with $K \in  G_-$ for a
pseudodifferential operator $Q$ such that $Q_+ = D$, is not unique: $K$ can be
multiplied from the right by pseudodifferential operators from $G_-$ with
constant coefficients (so as to commute with $D$). This arbitrariness has no
effect on the evolution equations (see 3.4), but does change the terms
$KxD^{n+1}K^{-1}$ as
$$
KxD^{n+1}K^{-1} \mapsto  KxD^{n+1}K^{-1} + \sum _{m\leq -1}h_m
D^{m+n}
$$
with constant $h_m$. This evidently can be viewed as a redefinition of $p$,
$$
p \mapsto  xD + \sum  h_m D^m,
$$
which is precisely the arbitrariness discussed in 2.9.2: $p$ is determined
unambiguously only as an element of
${\rm Der}{ \fr g}/{\rm ad}_{{\cal Z}_\Lambda }$. We will
stay with the above choice for $p$ with all $h_m = 0$.

{\bf {\it 3.7}.} As to the Virasoro {\it constraints}, they take a very simple
form ${ \fr L}_n = 0$ for $n \geq  0$. Choosing for simplicity $q = J$, a
generating expression for the constraints may be written as
$$
(K(P + \ell J) e^{D\ell } K^{-1})_- = 0
$$
where $\ell $ is a parameter and
$$
P = \sum _{r\geq 1}rt_r D^{r-1} = x + 2t_2D + \ldots
$$
Removing $(\ldots )_-$ results in replacing zero on the {\sc rhs} with a {\it
differential} operator $S(x,\ell )$:
$$
K(x)\circ (x + \ell J + \sum _{s\geq
2}st_sD^{s-1}) = S\circ K(x + \ell )
$$
This $S$ parametrizes independent degrees of freedom and therefore
`solves' the constraint. By virtue of the KP equations $S$
satisfies a set of non-local evolution equations \cite{[S2]}\cite{[S5]}
$$
{\partial S\over \partial t_r} = Q(x)^r_+S - SQ(x + \ell )^r_+
$$
Evolution equations of this type were studied also in \cite{[DLOPPS]}. It was
suggested there to look at them as the result of ``quantizing'' the spectral
parameter of standard, local, integrable equations. A very similar philosophy
was advocated in \cite{[Mo]} (``quantum'' Riemann surfaces)
also in relation with the Virasoro constraints.

{\bf {\it 3.7.1.}} {\sc Remark}. The Virasoro constraints on the KP hierarchy
give rise to higher constraints, related to the underlying algebra $W_\infty$
 \cite{[S5]}, which can be written in the form of a generating expression
depending on two parameters $\ell $ and $z$. For $J = 0$ one finds :
$$
\exp \left( \sum_{r\geq 2}t_r\left( \left(
{\partial \over \partial \ell } +
z\right)^r - {\partial^r\over \partial \ell^r}\right) \right)
(Ke^{zx}e^{\ell D}K^{-1})_- = (Ke^{\ell D}K^{-1})_-
$$

{\large {\bf $N$-KdV}}

\vspace{0.3cm}
{\bf {\it 3.8.}} Virasoro generators for the generalized KdV hierarchies follow
by a reduction \cite{[DDKM]} from the KP ones. Only the generators with mode
numbers $n = Nj$, $j \in  {\open Z}$ are compatible with the constraint
$$
(KD^NK^{-1})_-^{} = 0 $$
defining the $N-$reduction, and thus \cite{[S4]},
$$
{ \fr L}^{\rm KdV}_j = {1\over N} (K\sum^{N-1}_{\alpha =1}\sum
_{i\geq 0}(Ni + \alpha )t_{\alpha ,i} D^{N(i+j)+\alpha } K^{-1})_-
$$
where $t_{\alpha ,i} = t_{Ni+\alpha }$ and $x$ was identified with the time
$t_{0,1}$. As in the KP case, these ${\fr L}$'s are used to define vector
fields
$$
\hat {\fr l}^{{\rm KdV}}_j.f(Q) = {d\over d\epsilon } f(Q - \epsilon [{ \fr
L}^{\rm KdV}_j , Q] ) |_{\epsilon =0}\qquad   Q \equiv
KDK^{-1}.
$$
Strictly speaking, one should consider functions of the $N$-KdV Lax
operator $L = KD^NK^{-1}$ (which is differential by virtue of the above
constraint), rather than $Q = KDK^{-1}$. This and similar other
book-keeping style corrections are left to the reader.

{\bf {\it 3.9.}} Note that adding to $l_j^{\rm KdV}
= {1\over N} xD^{Nj+1}$ the piece $(Jj
+ q)D^{Nj}$, as suggested in 2.8, would not change the corresponding vector
fields $\hat l^{{\rm KdV}}_j$; for $j \geq  0$ this can be seen already for the
operators ${ \fr L}^{{\rm KdV}}_j$, while for $j < 0$ the $(\ldots
)_-^{}$-projection becomes irrelevant, so removing it one is left with
$KD^{Nj}K^{-1}$ which commutes with the Lax operator. On the other hand, $p =
{1\over N} xD$, being a representative of a class from ${\rm Der}{ \fr g}/{\rm
ad}_{{\cal Z}_\Lambda }$, can be changed by arbitrary powers of $D$, as
discussed in 3.6. The commutative subalgebra ${\cal Z}_\Lambda $ is the same as
in the KP case, $i.e.$, the one generated by $D^n$, $n \in  {\open Z}$.

{\bf {\it 3.10}}. {\sc Remarks}.

{\bf 1.} One can make contact with a more familiar formalism that has been used
to describe the Virasoro action on the KdV hierarchies, the tau function
approach. As shown in \cite{[S5]}, by evaluating the residue of the above
${ \fr L}^{\rm KdV}_j$, one finds the Virasoro generators
$$
{\Sc L}_j = {1\over N}
{1\over 2}\sum ^{N-1}_{\alpha =1}\sum ^{j-1}_{i=0}{\partial ^2\over \partial t_
{\alpha ,i}\partial t_{N-\alpha ,j-i-1}} + {1\over N}
\sum ^{N-1}_{\alpha =1}\sum _{i\geq 0}(Ni + \alpha )t_{\alpha ,i}
{\partial \over \partial t_{\alpha ,i+j}},\qquad   j > 0
$$
(and similar expressions for ${\Sc L}_{j\leq 0})$ which act on the tau function
of
the $N$-KdV hierarchy.

{\bf 2.} Imposing the Virasoro constraints ${ \fr L}^{\rm KdV}_j = 0$ for $j
\geq  -1$ in the case of the $N$-KdV hierarchy generates a system of
`higher' constraints as explained in \cite{[S5]}; these are summarized by the
equation $$ \left( Ke^{zPD^{1-N}} e^{\ell D^N} K^{-1}\right)_- = 0
$$ where $z$ and $\ell $ are parameters and $P$ is the same as in the KP case,
but with the times $t_{Nj}$ dropped.

\vspace{0.3cm}

{\large {\bf Toda}}

\vspace{0.3cm}
{\bf {\it 3.11}}. For the Toda lattice hierarchy \cite{[UT],[T]} the Lie
algebra ${\fr g} = { \fr gl}(\infty )$ is that of $\infty  \times
\infty$  matrices whose matrix elements will be labelled according to
$$
\Phi  = \sum _{i\in {\open Z}} \sum _{s\in {\open Z}} \phi _i(s) | s\rangle
\langle s| \Lambda ^i
$$
where $| s \rangle $ with $s \in  {\open Z}$ are the elements of an
orthonormal basis and
$$
\Lambda | s\rangle  = | s-1\rangle
$$

{\bf {\it 3.12}}. The $r-$matrix is chosen to be the one associated to the
decomposition of ${\fr gl}(\infty )$ into a sum of two
subalgebras which consist of (strictly) lower-triangular and (non-strictly)
upper-triangular matrices respectively. This can also be expressed as
$$
\Phi_+ = \sum_{i\geq 0} \sum_{s\in {\open Z}} \phi_i(s) |s\rangle
\langle s| \Lambda^i
$$
The dressing operator, denoted now as $W$,
therefore has the form
$$
W = 1 + \sum _{i\geq 1} \sum _{s\in {\open Z}} \phi _i(s)
| s\rangle \langle s| \Lambda ^{-i}
$$

{\bf {\it3.13}}. Matrix trace allows us to identify ${ \fr g}^\ast _-$ with the
strictly upper-triangular matrices. (We deliberately avoid the discussion, in
this illustrative example, of the conditions for the trace to exist.) The
hierarchy equations read,
$$
{\partial L\over \partial x_n} = - [(L^n)^{}_-, L ],\qquad
L = W\Lambda W^{-1},\quad  n \geq  1.
$$
where $x_n$ are the
hierarchy times (the $x$-times of the "two-dimensional" Toda lattice hierarchy
\cite{[UT]}).

{\bf {\it 3.14}}. Now one can construct the `bare' Virasoro generators in the
spirit of 2.9 by first defining a matrix $p$ by
$$
p| s\rangle  = s| s\rangle ,
$$
whence
$$
[\Lambda ,p] = \Lambda ,
$$
and then setting
$$
l_n = p\Lambda ^n,\qquad     A_n = \Lambda ^n
$$
and
$$
{ \fr l}^{\rm Toda}_n(J,q) = p\Lambda ^n + Jn\Lambda ^n + q\Lambda ^n +
\sum _{r\geq 1}rx_r \Lambda ^{n+r}.
$$
The ${\fr g}_+\oplus  { \fr g}_-$ decomposition can be
extended by setting $(p)_+ = p$; then
$$
{ \fr L}^{{\rm Toda}}_n(J,q) = \left( W
\left\{ [Jn + q + p ] \Lambda ^n + \sum _{r\geq 1}rx_r \Lambda ^{n+r}
\right\} W^{-1}
\right)_{\! -}
$$
These Virasoro generators were first derived in \cite{[S2]} via an
explicit calculation starting from the standard expression for the Virasoro
generators acting on the Toda tau function.

{\bf {\it 3.15}}. {\sc Remark}. The set of the Virasoro constraints ${ \fr
L}^{{\rm Toda}}_n(J,J) = 0$, $n \geq  0$, have been shown in \cite{[S3]} to
undergo a scaling into the KP Virasoro constraints ${ \fr L}^{\rm KP}_p(J,J) =
0$, $p \geq  -1$. By scaling we mean blowing up the asymptotic region $s \gg
1$
so that $s$ can be written as a formal expansion in inverse powers of
$\epsilon$
for $\epsilon \rightarrow  0$.  Then the scaling involves certain
$\epsilon$-dependent ansatze for the hierarchy times $x_m$ (expressing them
through new parameters $t_r$ which become KP-times) and for matrix elements of
$W$ (expressing them through what becomes coefficient functions of a KP
dressing
operator $K$).

\section{Matrix versus scalar $N$-KdV Virasoro constraints
and string equations}

\leavevmode\hbox to\parindent{\hfill}
The $N$-KdV hierarchies admit a very important formulation, which is
alternative to that used in Sect.3, in terms of {\it first} order differential
operators with matrix coefficients \cite{[DS]}. This formulation transpires
both the
geometric nature of the equations (hamiltonian reduction) and the Lie-
(in fact, Kac-Moody $sl(N))$ algebraic structure, which allows
generalizations to Kac-Moody algebras other than $\tilde {sl}(N)$. Specializing
the general construction of Sect.2 to the Drinfeld-Sokolov formulation of
$N$-KdV should therefore be viewed as an example of what this
construction becomes for the generalized KdV's of Drinfeld and Sokolov. The
respective Virasoro {\it constraints} would then correspond, in particular, to
gravity-coupled minimal models from the {\it ADE} classification of the
latter. (Note the paper \cite{[Y]} which deals with the {\it scalar}
formulation of
these.)

\vspace{0.3cm}
Thus, the $N$-KdV hierarchies that we have met in Sect.3
will now be called the $sl(N)$-KdV hierarchies and will be described in
the formalism which explicitly relies on the underlying Kac-Moody $sl(N)$
structure.

{\bf {\it 4.1}.} Recall that the $sl(N)$-{\rm KdV} hierarchy is formulated in
terms of a differential operator
$$
{\cal L} = {\partial \over \partial x} + {\bf L}
$$
where ${\bf L}$ is of the form
$$
{\bf L}(\zeta ) = \Lambda_N(\zeta ) +
\left(
\begin{array}{llllll}
\ast   &{}      &{}   &{}       &{}         &{} \\
{}  &\ast       &{}    &{}       &{}         &{}  \\
{}  &{}         &\ast &{}     &\mbox{0}   &{}  \\
{}  &{}         &{}   &\cdot     &{}       &{}  \\
{} &{\ast} &{}   &{}        &\cdot    &{}   \\
{}  &{}         &{}    &{}      &{}        &\ast
\end{array}
\right)
$$
with $\zeta $ being a spectral parameter and
$$
\Lambda _N(\zeta ) =
\left(
\begin{array}{llllll}
0   &1   &{}   &{}       &{}         &{} \\
{}  &0   &1    &{}       &{}         &{}  \\
{}  &{}  &\cdot &\cdot     &\mbox{0}   &{}  \\
{}  &{}  &{}   &\cdot     &\cdot       &{}  \\
0   &{}  &{}   &{}       &\cdot       &1   \\
\zeta &0 &{}    &{}      &{}         &0
\end{array}
\right)
\in  sl(N) , \qquad
\Lambda _N(\zeta )^N = \zeta \cdot 1
$$

By a gauge transformation
$$
{\partial \over \partial x} + {\bf L} = W \left( {\partial \over \partial x} +
{\bf L}_0\right) W^{-1},
$$
with
$$
W = 1 + \sum _{j\geq 1}\sum ^{N-1}_{s=0}|s\rangle w_j(s)\langle s|
\Lambda _N(\zeta )^{-j},
$$
one can bring ${\bf L}_0$ to the form \cite{[DS]}
$$
{\bf L}_0 = \Lambda _N(\zeta ) + \sum _{i\geq 0}f_i \Lambda _N(\zeta )^{-i}
$$
where $f_i$ are scalar functions.

{\bf {\it 4.2.}} The hierarchy flows are induced on gauge equivalence classes
by
the equations
$$
{d{\cal L}\over dt_{\alpha ,i}} = - [(\zeta ^iA^\alpha )_-^{}, {\cal
L}] , \quad    i \geq  0 , \quad   \alpha  \in  \{1,
2,\ldots ,  N-1\}
$$
in which
$$
A = W \Lambda _N(\zeta ) W^{-1}
$$
and either of the two possibilities can be
chosen as the definition of $(\ldots )_-$ (and therefore of the $r$-matrix):
projection onto negative powers of $\zeta $ or onto negative powers of $\Lambda
_N(\zeta )$. In the latter case one uses the presentation
$$
{\cal N} = \sum _{i\leq n} \sum ^{N-1}_{s=0}|s\rangle n_i(s)\langle
s|\Lambda _N(\zeta )^i
$$
for an arbitrary matrix from the algebra; then
$$
{\cal N}_- = \sum _{i<0} \sum ^{N-1}_{s=0}|s\rangle n_i(s)\langle s|\Lambda
_N(\zeta )^i
$$
The two definitions lead to the same evolutions on the gauge equivalence
classes.

{\bf {\it 4.3.}} The Virasoro transformations of ${\cal L}$ are given by,
$$
\hat {\fr l}_j {\cal L} = [{\fr L}_j,  {\cal L} ]
$$
with
$$
{\fr L}_j = \left( W \left\lbrace  - \zeta  {\partial \over \partial \zeta } +
\hat \pi  + \sum ^{N-1}_{\alpha =1}\left(  \sum _{i\geq 0}(i + {\alpha \over N}
)t_{\alpha ,i} \zeta ^i \right)  \Lambda _N(\zeta )^\alpha  \right\rbrace
\zeta ^j W^{-1} \right)_{\! -}
$$
where
$$
\hat \pi  = {1\over N}
\left(
\begin{array}{llllll}
1   &{}       &{}   &{}       &{}        &{} \\
{}  &2        &{}    &{}       &{}       &{}  \\
{}  &{}       &\cdot &{}       &{}       &{}  \\
{}  &{}       &{}   &\cdot     &{}       &{}  \\
{}  &{}        &{}   &{}        &\cdot   &{}   \\
{}  &{}       &{}    &{}      &{}        &N
\end{array}\right)
+ \sum    h_i \Lambda _N(\zeta )^{-i}
$$
and $h_i$ are arbitrary {\it scalar} functions. This is the same type of
arbitrariness as discussed in 3.6 and 3.9.

{\bf {\it 4.4.}} Now the Virasoro {\it constraints}  read
$$
{\fr L}_j = 0 \quad {\rm for }\quad j \geq  -1.
$$
In each of these equations we remove the $(\ldots )_-$-projection, which
results in replacing zero on the {\sc rhs} with a  $-S^{(j)}_+$ (with
$S^{(j)}_+$ being purely $(\ldots )_+-$matrices). Further, one readily
discovers
that
$$
S^{(j)}_+ = (j + 1)\zeta ^j + \zeta ^{j+1}S^{(-1)}_+
$$
and thus the Virasoro constraints lead to a single equation of the form
$$
{\partial W^{-1}\over \partial \zeta } = W^{-1} S_+ + \left( \hat \pi  + 1 +
\sum_{\alpha ,i}\left( i + {\alpha \over N}\right) t_{\alpha ,i} \zeta ^i
\Lambda _N(\zeta )^\alpha  \right)  \zeta ^{-1} W^{-1}
$$
where $S_+$ is an arbitrary $(\ldots )_+-$matrix. Due to obvious reasons, it
is advantageous to choose here the definition of
$(\ldots )_+$ as positive powers of the spectral parameter $\zeta$ .

{\it {\bf 4.4.1}}. {\sc Remark}. As an aside, note that for the matrix
$$
\Psi  = W\exp\sum_{\alpha ,i}t_{\alpha ,i} \zeta ^i \Lambda _N(\zeta )^\alpha
$$
the linear equation takes the form (cf.\cite{[Mo]})
$$
- {\partial \Psi \over \partial \zeta } = S_+\Psi  + \Psi (\hat \pi  +
1)\zeta ^{-1}
$$
Together with the standard linear problem for the $sl(N)$-KdV
hierarchy this gives a "generalized linear system" whose
integrability leads, in
particular, to the string equation.

\vspace{0.3cm}
Now we investigate the consistency between Virasoro generators in the scalar
and matrix formalisms for the KdV hierarchy.

\vspace{0.3cm}
{\bf {\it 4.5.}} Recall first the standard procedure to rederive the scalar
formalism from the matrix one \cite{[DS]}. Denote by ${\Sc H}$ a vector space
whose elements are columns of height $N$ consisting of formal Laurent series in
$\zeta $, which might also bear a dependence on $x$. For a fixed Lax operator
${\cal L}$ one can make ${\Sc H}$  into a module over the ring of
pseudodifferential operators by
$$
_{\dot{\cal L}}^{\phantom{Z}}:
 \psi {\rm Diff} \times  {\Sc H}  \rightarrow  {\Sc H},
$$
$$
F
_{\dot{\cal L}}^{\phantom{Z}}
\eta  = \sum ^N_{i=-\infty }f_i {\cal L}^i \eta ,
$$
where on the {\sc rhs} the action by the usual matrix multiplication and
application
of $D = \partial /\partial x$ are understood. The definition is correct since
$$
D_{\dot{\cal L}}^{\phantom{Z}}(f \eta )
 = (fD)\eta  + (\partial f) \eta  = (D \circ  f)
_{\dot{\cal L}}^{\phantom{Z}}
\eta
$$
for a scalar function $f(x)$, whence
$$
FG
_{\dot{\cal L}}^{\phantom{Z}}
\eta  = F
_{\dot{\cal L}}^{\phantom{Z}}
(G
_{\dot{\cal L}}^{\phantom{Z}}
\eta ) \quad {\rm for} \quad F,G \in  \psi {\rm Diff}.
$$

{\bf {\it 4.6.}} Now, the scalar Lax operator can be recovered starting from
the
matrix one from the relation
$$
L
_{\dot{\cal L}}^{\phantom{Z}}
\eta _0
 = \zeta \eta _0\qquad     \eta _0 \equiv
  \left(
\begin{array}{l}
0\\
0\\
\cdot\\
\cdot\\
\cdot\\
1
\end{array}
\right) \hbox{ .}
\eqno{(\ast)}
$$
It turns out that this defines $L$ uniquely, and it has the form
$$
L = D^N + \sum ^{N-1}_{i=0}u_i D^i,
$$
where $u_i$ are certain scalar functions of $x$. It is now a result of the
standard theory that this operator can be represented as
$$
L = KD^NK^{-1}
$$
where $K$ is of the same form as in 3.2.

\vspace{0.3cm}
We want to see whether the above relation between the matrix and scalar Lax
operators is preserved by the Virasoro action. That is, viewing the {\sc lhs}
of
$(\ast )$ as a function  $\{L\}: {\cal L} \mapsto  \{L\}({\cal L})$, we deform
both the function and the argument, by
$$
\hat {\fr l}^{{\rm sc}}_j.L = [{\fr L}^{{\rm sc}}_j , L]
$$
and
$$
(\hat {\fr l}^{\rm m}_j\{L\})({\cal L}) = {d\over d\epsilon } \{L\}({\cal L} -
\epsilon
[{\fr L}^{\rm m}_j,{\cal L} ] )
\mid_{\epsilon =0}
$$
respectively. Here ${\fr L}^{\rm sc}_j$ and ${\fr L}^{\rm m}_j$ are borrowed
from 3.8 and 4.3 respectively.

It turns out that this combined action of scalar and matrix Virasoro generators
does {\it not} leave $(\ast )$ invariant. Instead, we have

\vspace{0.3cm}
{\bf 4.7.} {\sc Theorem} {\sl ({\underline {consistency relation between scalar
and matrix Virasoro generators}}). For the above Virasoro generators it follows
$$
\delta ^{\rm tot}_j(L _{\dot{\cal L}}^{\phantom{Z}} \eta _0 - \zeta \eta _0) =
0,
$$
where}
$$
\new
\begin{array}{rcl} \delta^{{\rm tot}}_j {\cal L} &=& - [{ \fr L}^{\rm m}_j
, {\cal L}]\\ \delta^{{\rm tot}}_j   L      &=& - [{ \fr L}^{\rm sc}_j , L]\\
\delta^{{\rm tot}}_j \zeta    &=& \zeta^{j+1} \end{array}
$$

\vspace{0.3cm}

{\sc Remark} (trivial). Note that the bottom line is just the action of
reparametrizations $\zeta ^{j+1} {\partial \over \partial \zeta }$ on a complex
parameter.

\vspace{0.3cm}
{\sc Proof}. We evaluate directly the action of $\delta ^{\rm tot}_j$ on $(L
_{\dot{\cal L}}^{\phantom{Z}}
\eta _0):$
$$
\delta ^{\rm tot}_j(L
_{\dot{\cal L}}^{\phantom{Z}}
\eta _0) = ({\fr L}^{\rm sc}_j L)
_{\dot{\cal L}}^{\phantom{Z}}
\eta _0 - (L {\fr L}^{\rm sc}_j )
_{\dot{\cal L}}^{\phantom{Z}}
\eta _0 - \left\lbrace  [{\fr L}^{\rm m}_j\hbox{ , } {\cal L}^N ] + \sum
^{N-1}_{i=0}u_i [{\fr L}^{\rm m}_j,{\cal L}^i ] \right\rbrace \eta _0,
$$
where $u_i$ are the above coefficients of the unperturbed scalar Lax operator
$L$. Taking into account that  $[{\fr L}^{\rm m}_j$ , $u_i ] = 0$  and using
again equation $(\ast )$ of 4.6 (and also the ring homomorphism property 4.5),
we bring the last equation to the form,
$$
\new
\begin{array}{rcl}
\delta^{\rm tot}_j
\left(    L_{ \dot{\cal L} }^      { \phantom{Z}  }        \eta _0\right)
 &=&
{{\fr L}^{\rm sc}_j }{}_{\dot{{\cal L}}}^{ \phantom{Z}}
\zeta \eta _0 - {\fr L}^{\rm m}_j \zeta \eta _{0} - L
_{ \dot{\cal L} }^{ \phantom{Z}  }
       \left(
 {\fr L}^{\rm sc}_j   {}_{ \dot{\cal L} }^{ \phantom{Z}  }
               \eta _0 - {\fr L}^{\rm m}_j \eta _0
        \right) \\
{}&=& \zeta^{j+1} \eta _0 + (\zeta  - L
_{ \dot{\cal L} }^{ \phantom{Z}  })
      \left\{
                  {{\fr L}^{\rm sc}_j}
                  _{\dot{\cal L}}^{\phantom{Z}}
                   \eta _0 - {\fr L}^{\rm m}_j \eta _0
       \right\}.
\end{array}
\eqno{(\ast \ast) }
$$
Next, the scalar generator ${\fr L}^{{\rm sc}}_j$, which is of course nothing
but ${ \fr L}^{{\rm KdV}}_j$ we had in 3.8, can be rewritten as,
$$
{\fr L}^{\rm sc}_j = {1\over N} ( KxD^{Nj+1} K^{-1} )_-^{}
+ \sum ^{N-1}_{\alpha
=1}\sum _{i\geq 0} \left( i + {\alpha \over N}\right)
t_{\alpha ,i} \left(  L^{i+j+{\alpha
\over N}} \right)^{}_-
$$
(Recall that $L = KD^NK^{-1}$.)

Now, as shown in \cite{[DS]},
$$
{\left(  {L^{i+j+{\alpha \over N}}} \right)^{}_-}
_{ \dot{\cal L} }^{ \phantom{Z}  }
\eta_0 = (W \Lambda _N(\zeta )^{N(i+j)+\alpha } W^{-1})_-^{}\eta_0.
$$
where $W$ is defined just by the relation of 4.1 and $(\ldots )_-$ on the {\sc
rhs}
refers to negative powers of $\Lambda _N(\zeta )$. Therefore, it follows for
the $\sum\sum $-piece of ${\fr L}^{{\rm sc}}_j$ that
$$
\sum ^{N-1}_{\alpha =1}\sum_{i\geq 0}\left( \ldots \right)
_{ \dot{\cal L} }^{ \phantom{Z}  }
\eta _0 = \left( W \left(  \sum _{\alpha ,i}\left( i + {\alpha \over N}\right)
t_{\alpha ,i}
\zeta ^i \Lambda _N(\zeta )^\alpha  \right)  \zeta ^j W^{-1}
\right)_{\! -}\eta _0,
$$
which coincides with the respective piece in the expression for ${\fr L}^{\rm
m}_j$ from 4.3.  It therefore remains to calculate the  $ _{ \dot{\cal L} }^{
\phantom{Z}  } $ -action of the first term in ${\fr L}^{{\rm sc}}_j.$

As shown in \cite{[DS]}, the $(\ldots )_-$'s are ``equivariant''. Thus it
suffices to
determine an ${\cal X}_j$ from the formula
$$
{1\over N} (KxD^{Nj+1} K^{-1})
_{ \dot{\cal L} }^{ \phantom{Z}  }
\eta _0 = {\cal X}_j \eta _0
$$
The pseudodifferential operator from the {\sc lhs} satisfies the following
commutation relation:
$$
\left[ L^{{1\over N}}\hbox{ , } {1\over N} (KxD^{Nj+1} K^{-1}) \right]  =
{1\over N} L^{j+{1\over N}}.
$$
Therefore for ${\cal X}_j$ one should have
$$
\left[ W \Lambda _N(\zeta ) W^{-1}, {\cal X}_j \right]  = {1\over N}
W\ \Lambda _N(\zeta )^{Nj+1} W^{-1},
$$
which determines ${\cal X}_j$ in the form
$$
{\cal X}_j = W \biggl(  - \zeta  {\partial \over \partial \zeta } + {1\over N}
\left(
\begin{array}{llllll}
1   &{}       &{}   &{}       &{}         &{} \\
{}  &2        &{}    &{}       &{}         &{}  \\
{}  &{}       &\cdot &{}     &\mbox{0}   &{}  \\
{}  &{}       &{}   &\cdot     &{}       &{}  \\
{} &\mbox{0}  &{}   &{}        &\cdot    &{}   \\
{}  &{}       &{}    &{}      &{}        &N
\end{array}\right)
+ \sum    h_i
\Lambda _N(\zeta )^{-i}\biggr)  \Lambda _N(\zeta )^{Nj}\ W^{-1}
$$
with $h_i$ arbitrary scalars as in the expression for $\hat \pi $ in 4.3.
Putting everything together, we thus arrive at
$$
%\new
%\begin{array}{l}
{L^{{\rm sc}}_j}
_{\dot{\cal L}}^{\phantom{Z}}
\eta_0
= \biggl( W \biggl( - \zeta  {\partial \over \partial \zeta } + {1\over N}
\left(
\begin{array}{llllll}
1   &{}       &{}   &{}       &{}         &{} \\
{}  &2        &{}    &{}       &{}         &{}  \\
{}  &{}       &\cdot &{}     &\mbox{0}   &{}  \\
{}  &{}       &{}   &\cdot     &{}       &{}  \\
{} &\mbox{0}  &{}   &{}        &\cdot    &{}   \\
{}  &{}       &{}    &{}      &{}        &N
\end{array}\right)
%\\
+ \sum _{\alpha ,i}\left( i+
{\alpha \over N}\right) t_{\alpha ,i} \zeta ^i \Lambda_N(\zeta )^\alpha  +
\ldots \biggr)\zeta ^j W^{-1}\biggr)_{\! -}\eta _0
%\end{array}
$$
where the ellipsis denotes the $h-$terms. These can be chosen to coincide
with the corresponding ones from the formula in 4.3
\footnote{ and, moreover, put to zero throughout.
}
and thus the curly bracket in $(\ast \ast )$ vanishes, which completes the
proof.
$\Box$

{\bf {\it 4.8.}} An important difference between the scalar and matrix
formulations has to do with the string equation. We have seen that the Virasoro
generators and therefore the Virasoro constraints in the two cases agree.
However, although the string equation is a consequence of Virasoro constraints,
its form in the matrix formalism is not a naive replica of the scalar one. We
have seen in 3.8 that, in particular, ${ \fr L}^{\rm KdV}_{-1}$, which we now
denote as ${ \fr L}^{\rm sc}_{-1}$, is of the form ${ \fr L}^{{\rm sc}}_{-1}
= ( {{\fr P}} ^{{\rm sc}})_-$, so that the $(-1)$-Virasoro constraint takes the
form of the condition that $   {{\fr P}}^{\rm sc}$ be a {\it differential}
operator. As such, it satisfies the string equation
$$
[L, {{\fr P}} ^{{\rm
sc}}] = 1
$$
where $L = KD^NK^{-1}$ is the Lax operator.  On the other hand we
have seen that the matrix counterpart of this equation reads
$$
[\zeta ,
{\fr P}^{\rm m}] = 1,
$$
where, similarly to the scalar case,
$({\fr P}^{\rm m})_- = {\fr L}^{\rm m}_{-1}$,
whereas, generally, $[{\cal L}, {\fr P}^{\rm m}] \neq 1$.  It should be
noted also that, contrary to the scalar case, $\zeta $ is {\it not} an
eigenvalue of the Lax operator ${\cal L}$ in the matrix case.

A similar remark would apply also to similar constructions for other
Kac-Moody algebras.

\section{Concluding remarks}

\leavevmode\hbox to\parindent{\hfill}
Symmetries of (constrained) integrable hierarchies, `higher' than the Virasoro
ones, have been a subject of  considerable interest recently
\cite{[AS],[FKN1],[FKN2],[FKN3],[G],[IM2],[S5]}. In the formalism we have
developed in
this paper, the higher symmetries can be generated from the Virasoro ones by
essentially algebraic manipulations. Consider, for instance,
Virasoro-constrained hierarchies as in 2.10 and assume for simplicity
that ${1\over 2} (1 - R)$ is a projection (as was the case in all the examples
in Sects. 3 and 4). Then,
$$
\left( {\rm Ad}_K .({ \fr l}_i { \fr l}_j)\right) _-
= \left( ({\rm Ad}_K .{ \fr l}_i ) ( {\rm Ad}_K .{ \fr l}_j
)\right) _-
$$
which is again zero by the Virasoro constraint saying that each factor is a
pure $(\ldots )_-$. Thus ${ \fr l}_i { \fr l}_j$ and, similarly, higher order
monomials, give
us an infinite set of `higher' constraints whose algebra is not difficult to
derive. Note that one naturally uses in this analysis the {\it associative}
structure in ${\rm End}{\fr g}$ to compose derivations and products thereof. It
is such an `associative mechanism' that allowed one to derive, for instance,
the
constraints of 3.7.1 in the KP case. The interested reader would formulate the
necessary requirements that the $r$-matrix must satisfy in order that the
construction make sense rigorously.

\vspace{1cm}
{\bf {\sc Acknowledgements.}}
It is a pleasure to acknowledge a kind hospitality
extended to me at the Nor-Amberd (Yerevan) workshop where this paper was
written. I especially thank R.Flume for encouragement.

              \end{document}